\documentclass[sigconf]{acmart}
\copyrightyear{2026}
\acmYear{2026}
\setcopyright{cc}
\setcctype{by}
\acmConference[SIGIR '26]{Proceedings of the 49th International ACM SIGIR Conference on Research and Development in Information Retrieval}{July 20--24, 2026}{Melbourne, VIC, Australia}
\acmBooktitle{Proceedings of the 49th International ACM SIGIR Conference on Research and Development in Information Retrieval (SIGIR '26), July 20--24, 2026, Melbourne, VIC, Australia}
\acmDOI{10.1145/3805712.3809692}
\acmISBN{979-8-4007-2599-9/2026/07}

\usepackage{eso-pic}
\usepackage{multirow} % For \multirow
\usepackage{arydshln} % <--- 导入这个宏包来画虚线
\usepackage{booktabs} % 常用，提供更美观的表格线，但在这里我们用 hline/hdashline
\usepackage{amsmath}
\usepackage{makecell}
\usepackage{xcolor}
\usepackage[table]{xcolor}
\usepackage{enumitem}
\usepackage[dvipsnames]{xcolor}
\usepackage{graphicx}
\usepackage{booktabs}
\usepackage{multirow}
\usepackage{makecell}
\usepackage{algorithm}
\usepackage{algorithmic}
\usepackage{geometry}
\usepackage{amsmath} % 用于数学符号
\usepackage[export]{adjustbox}
\usepackage[skins]{tcolorbox}
\usepackage[normalem]{ulem}

\AtBeginDocument{%
  }

\newcommand{\LongCatHeader}{%
  \AddToShipoutPictureFG*{%
    \AtPageUpperLeft{%
      \raisebox{-1.25cm}[0pt][0pt]{%
        \hspace{1.9cm}%
        \begin{minipage}{\dimexpr\paperwidth-3.8cm\relax}
        \includegraphics[width=0.15\textwidth]{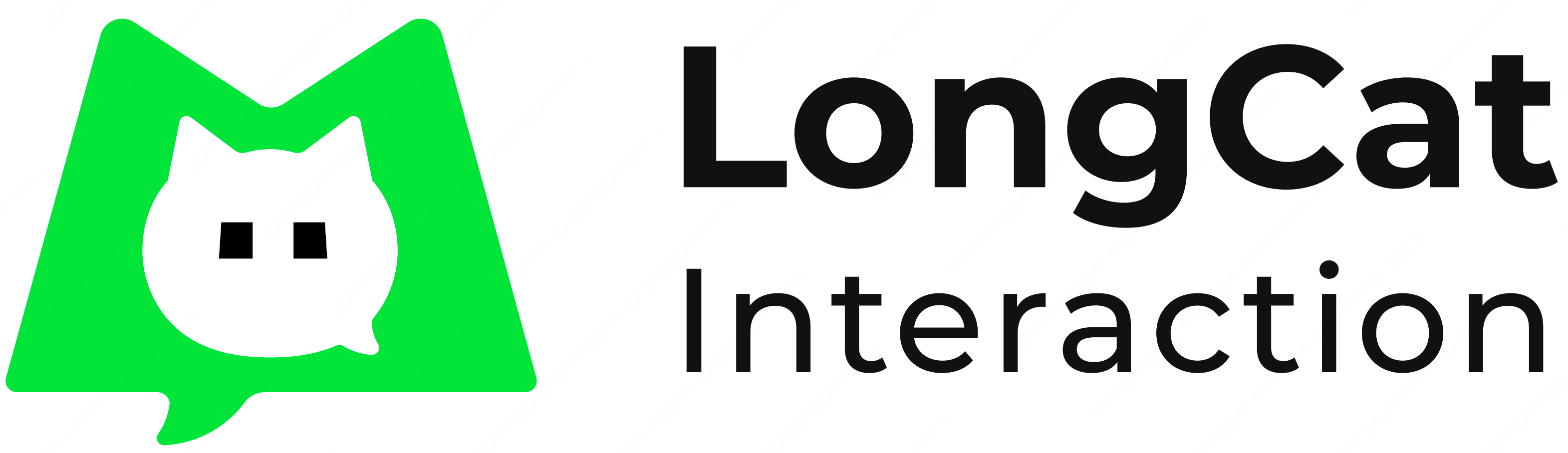}\par
          \rule{\linewidth}{0.3pt}
        \end{minipage}%
      }%
    }%
  }%
}

\title{Purifying Multimodal Retrieval: Fragment-Level Evidence \\ Selection for RAG}

\author{Xihang Wang}
\authornote{These authors contributed equally to this work.}
\affiliation{%
  \institution{Zhejiang University}
  \city{Hangzhou}
  \country{China}}
\email{xihangwang@163.com}

\author{Zihan Wang}
\authornotemark[1] % 和上面共享同一个标注
\affiliation{%
 \institution{Meituan LongCat Interaction Team}
 \city{Beijing}
 \country{China}}
\email{wangzihan14@meituan.com}

\author{Chengkai Huang}
\affiliation{%
  \institution{University of New South Wales,}
  \institution{Macquarie University}
  \city{Sydney}
  \country{Australia}}
\email{chengkai.huang1@unsw.edu.au}
\authornote{Corresponding author.}

\author{Cao Liu}
\affiliation{%
 \institution{Meituan LongCat Interaction Team}
 \city{Beijing}
 \country{China}}
\email{liucao@meituan.com}
 
\author{Ke Zeng}
\affiliation{%
 \institution{Meituan LongCat Interaction Team}
 \city{Beijing}
 \country{China}}
\email{zengke02@meituan.com}

\author{Quan Z. Sheng }
\affiliation{%
  \institution{Macquarie University}
    \city{Sydney}
  \country{Australia}}
\email{michael.sheng@mq.edu.au}

\author{Lina Yao}
\affiliation{%
  \institution{University of New South Wales,}
  \institution{CSIRO’s Data61}
    \city{Sydney}
  \country{Australia}}
\email{lina.yao@unsw.edu.au}

\renewcommand{\shortauthors}{Xihang et al.}

\begin{abstract}
Multimodal Retrieval-Augmented Generation (MRAG) is widely adopted for Multimodal Large Language Models (MLLMs) with external evidence to reduce hallucinations. Despite its success, most existing MRAG frameworks treat retrieved evidence as indivisible documents, implicitly assuming that all content within a document is equally informative. In practice, however, sometimes only a small fraction of a document is relevant to a given query, while the remaining content introduces substantial noise that may lead to performance degradation.
We address this fundamental limitation by reframing MRAG as a fine-grained evidence selection problem. We propose Fragment-level Evidence Selection for RAG (FES-RAG), a framework that selects atomic multimodal fragments rather than entire documents as grounding evidence. FES-RAG decomposes retrieved multimodal documents into sentence-level textual fragments and region-level visual fragments, enabling precise identification of evidence that directly supports generation.
To guide fragment selection, we introduce Fragment Information Gain (FIG), a principled metric that measures the marginal contribution of each fragment to the MLLM’s generation confidence. Based on FIG, we distill fragment-level utility judgments from a high-capacity MLLM into a lightweight selector, achieving accurate evidence selection with low inference overhead.
Experiments on the $M^2$RAG benchmark show that FES-RAG consistently outperforms state-of-the-art document-level MRAG methods, achieving up to 27\% relative improvement in CIDEr. By selecting fewer yet more informative fragments, our approach substantially reduces context length while improving factual accuracy and generation coherence.

\end{abstract}

\ccsdesc[500]{Information systems~Information retrieval}

\keywords{Retrieval-Augmented Generation, Large Language Model, Multi-modal Large Language Model}

\begin{document}

\LongCatHeader
\maketitle

\section{Introduction}

\begin{figure}
    \centering
\includegraphics[width=\linewidth]{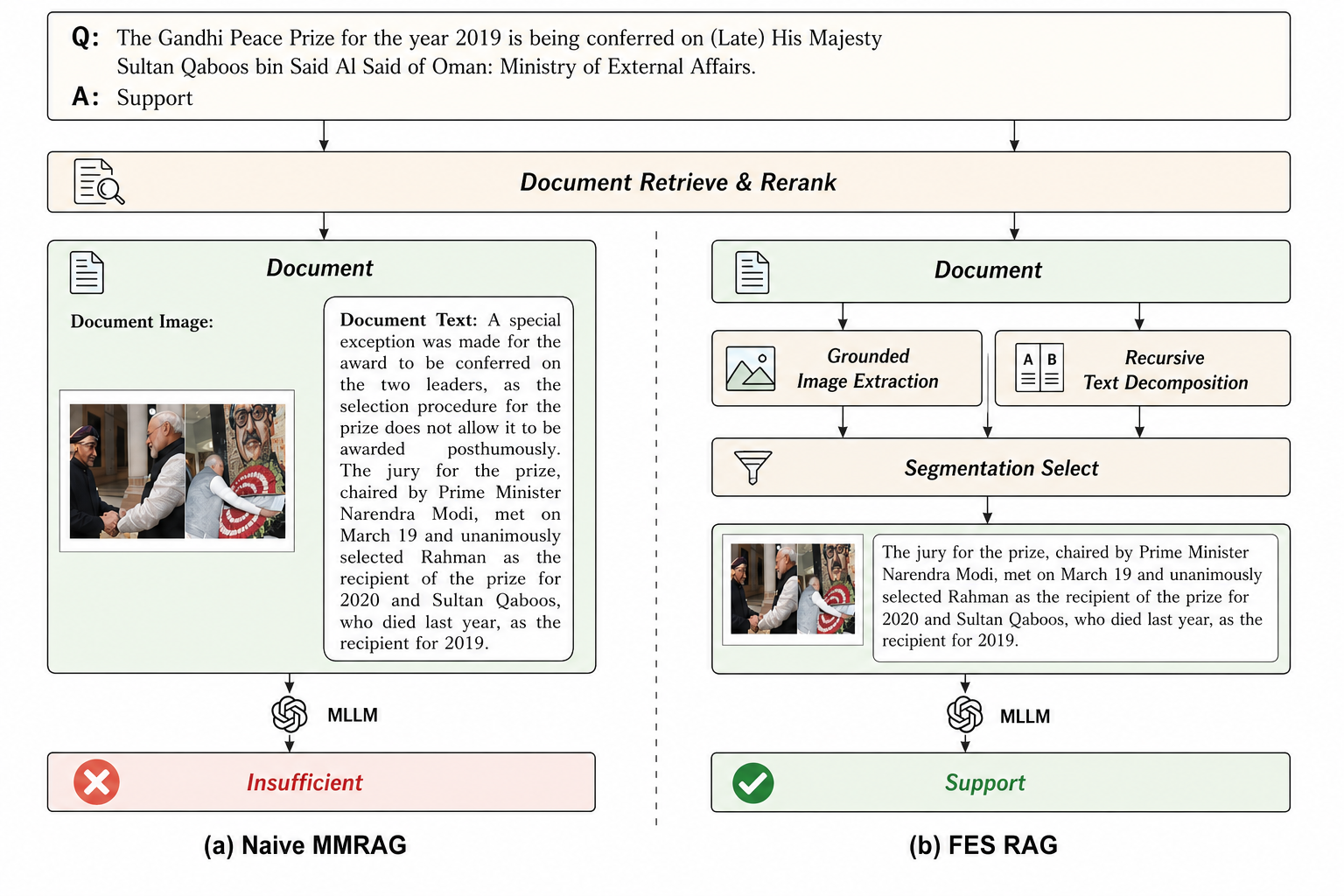}
\vspace{-1em}
    \caption{Comparison of the traditional MRAG pipeline (a) and our FES-RAG framework (b). Traditional methods only perform document-level retrieval and reranking, while our approach adds an additional document segmentation and fragment selection stage to purify the evidence.}
    \label{fig:1}
\end{figure}
% \begin{figure}
%     \centering
%     \includegraphics[width=1.1\linewidth]{Fig/intro_img_v2.png}
%     \caption{Comparison of the traditional MRAG pipeline and our FES-RAG framework. Traditional methods only perform document-level retrieval and reranking, while our approach adds an additional document segmentation and fragment selection stage to purify the evidence.}
%     \label{intro_img}
% \end{figure}

Multimodal Large Language Models (MLLMs) have shown remarkable capability in understanding and generating content across various multimodal tasks \cite{mei2025survey,huang2025towards,lou2025speechagent,gu2026learning}. 
Despite these advances, MLLMs are frequently constrained by inherent limitations such as hallucinations and outdated knowledge \cite{mei2025survey,wang2026scenealign}. To mitigate these challenges, Multimodal Retrieval-Augmented Generation (MRAG) has emerged as a critical paradigm \cite{mei2025survey}. 
By retrieving relevant external multimodal evidence, such as specific text documents or visual images, MRAG can enhance the accuracy and reliability of the MLLM's generation \cite{abootorabi2025ask,wang2026megrag}.

The effectiveness of MRAG hinges fundamentally on the quality of the retrieved context \cite{mragbench, re-vilm,jiao2026doctor}. When the retrieved context is contaminated with irrelevant, redundant, or contradictory information, MLLMs are prone to being misled by such distracting noise, which often precipitates factual hallucinations and undermines the reliability of the output \cite{Reliable_MM, rag-check, distracting}. This vulnerability is further compounded by the lost-in-the-middle effect, a phenomenon where MLLMs struggle to identify and leverage critical evidence when it is situated within the intermediate segments of a long context window \cite{liu2023lost,novel_middle,found_middle}. Consequently, even high-quality evidence can fail to support effective reasoning if it is obscured by surrounding noise or poorly positioned within the input.

To address this challenge, our research focuses on refining retrieved data at a granular level. The primary objective is to facilitate a transition from the monolithic documents toward atomic, fine-grained segments. We seek to investigate how the marginal utility of evidence can be quantified at the atomic level, thereby enabling the filtering of irrelevant, redundant, or contradictory noise that typically compromises downstream reasoning. Ultimately, our goal is to provide MLLMs with a compact context that contains only essential evidence, so that generation is guided by more reliable multimodal information.

% Consequently, our core research focus shifts to the problem of refined selection for effectively retrieved information: on the basis of initially retrieved relevant content, we aim to further screen and filter out irrelevant, redundant, or contradictory information that may interfere with subsequent reasoning and generation. This refined selection process is intended to purify the retrieved evidence pool and retain only the most valuable information that truly supports accurate multimodal reasoning.

%fragment 粒度、使用方式、形式是否有本质不同;说明一下document为什么会失败;
%引case study

Current relevant approaches, however, fall short of this aim due to their reliance on coarse-grained document processing \cite{infogain,selfrag}. Methods such as InfoGain-RAG \cite{infogain} and MM-R5 \cite{mm-r5} focus on document-level filtering and reranking, yet they still operate on the premise of treating retrieved content as a monolithic document. This limitation is further amplified in the multimodal domain: retrieved images often contain only a small Region of Interest (ROI) that is pertinent to the query, while the majority of the visual content consists of cluttered backgrounds, irrelevant objects, or distracting scenes \cite{object-mm}. As shown in Figure \ref{fig:1} (a), the presence of competing textual entities (e.g., the 2020 laureate) and irrelevant visual ROIs (e.g., unrelated political figures in the right part) distracts the model, leading it to mistakenly conclude that the evidence for the 2019 prize is insufficient.

%提出selector，提出segmentation
To address these limitations, we propose Fragment-level Evidence Selection for RAG (FES-RAG), a framework that moves beyond conventional document-level reranking toward fine-grained fragment purification. FES-RAG introduces a multimodal segmentation pipeline that decomposes retrieved documents into atomic fragments. To quantify the utility of these fragments, we extend the concept of Information Gain (IG) \cite{infogain,izacard2022atlas} from the document level to the fragment level, introducing Fragment Information Gain (FIG). This metric measures the marginal contribution of each atomic unit to the MLLM’s generation confidence, providing a high-fidelity supervision signal for training the fragment selector. To facilitate real-time deployment, we propose a knowledge distillation (KD) training strategy to bridge the gap between selection precision and inference efficiency. 

By feeding only high-information-density, purified evidence into the generator, we ensure that the MLLM can focus its reasoning attention exclusively on the fragments that directly support the answer generation process. As shown in Figure \ref{fig:1} (b), this decomposition enables the targeted extraction of critical information while filtering out irrelevant text sentences and visual regions. 

Our primary contributions are summarized as follows:
\begin{itemize}
    \item 
    % 提出新framework，对文档进行segmentation，以及对segment进行select
    We propose the FES-RAG framework, a novel paradigm shifting from document-level to fragment-level evidence processing for MRAG. It integrates systematic multimodal document segmentation and fine-grained fragment selection to purify retrieved evidence, effectively mitigating noise interference in traditional coarse-grained systems.
    \item 
    % 文档分割、FIG计算，得到高质量dataset
    We develop a scalable pipeline combining multimodal segmentation and FIG calculation, which enables precise quantification of fragment utility and provides high-quality supervision data for fine-grained evidence selection.
    \item 
    % 知识蒸馏
    We design a knowledge distillation-based training strategy for the evidence selector, transferring the fragment utility-judging capability of a high-capacity teacher MLLM to a lightweight student model, balancing evidence extraction precision and real-time inference efficiency.
    \item 
    % 实验效果     
    Extensive experiments on the M2RAG benchmark (MMQA, Image Captioning, Fact Verification) show our FES-RAG outperforms state-of-the-art coarse-grained baselines consistently, achieving up to 27\% relative CIDEr improvement while reducing context token consumption.
\end{itemize}

%（1）介绍 RAG 的基本概念：阐述通过检索外部知识来缓解大模型幻觉并更新时效性知识的有效性 。 （2）介绍 MRAG 的发展脉络：描述从简单的视觉对齐（如 CLIP）到复杂的“检索-生成”流水线（如 RA-VLM）的演进 。 （3）分析 MRAG 的颗粒度挑战：指出目前方法多停留于文档或图像等单体级别，容易导致“注意力稀释” 。 （4）介绍信息增益的理论来源：回顾香农信息论在特征选择和量化模型不确定性减少方面的基础作用 。 （5）介绍 RAG 中的置信度评价方法：讨论 Self-RAG 和 ActiveRAG 等通过模型内部状态评估证据实用性的尝试 。 （6）介绍文档级信息增益 (DIG)：说明 InfoGain-RAG 如何通过计算模型置信度变化来量化整个文档的贡献 。 （7）分析 DIG 在处理内部噪声时的局限：指出文档级度量往往会忽略长文档或复杂图像内部存在的低信噪比问题 。 （8）总结本研究从文档到片段的延伸动机：强调将信息增益延伸至原子片段层级（FIG）以实现精准证据提取的必要性 。

\section{Related Work}

\subsection{Multi-modal RAG}

The integration of external knowledge into Large Language Models (LLMs) through Retrieval-Augmented Generation (RAG) has proven highly effective in mitigating hallucinations and improving knowledge currency \cite{lewis2020rag, rag-survey,jiao2026prunerag,huang2025embedding}. Building upon the success of text-based RAG, MRAG extends this paradigm to MLLMs, enabling them to reason over cross-modal evidence such as images, videos, and structured tables \cite{multi-source, m2rag}. Early works in MRAG primarily focused on aligning visual features with textual queries using dual-encoder architectures like CLIP \cite{CLIP} or BLIP \cite{Junnan} to retrieve relevant images that complement textual prompts.

Recent advancements have shifted toward more complex retrieve-then-generate pipelines. For instance, frameworks like RoRA-VLM \cite{rora-vlm} and Re-ViLM \cite{re-vilm} demonstrate that retrieving pertinent visual context significantly improves performance on knowledge-intensive tasks such as multimodal question answering (MMQA) and image captioning. To benchmark these capabilities, the $M^2$RAG dataset was introduced, providing a rigorous evaluation of how MLLMs handle long-context multimodal retrieval across diverse real-world scenarios \cite{m2rag}. 

Most existing methods operate at the document level, treating a retrieved item as a monolithic unit \cite{mm-r5}. However, providing entire documents often introduces substantial noise, such as irrelevant background objects in images or irrelevant sentences in text, which can distract the MLLM and lead to degraded reasoning quality \cite{irrelevant_context,power_of_noise}. Our work addresses this gap by proposing a fine-grained selection mechanism based on information gain, ensuring that only information-dense evidence is presented to the generator. Notably, to the best of our knowledge, there are currently no existing baselines that implement sentence-level reranking specifically within the MRAG domain.

\subsection{Information Gain}
The concept of information gain, rooted in Shannon's information theory, has traditionally served as a cornerstone for feature selection and decision tree induction by quantifying the expected reduction in entropy after observing a feature~\cite{IG_tree}.
In the context of RAG, the paradigm has shifted from measuring simple semantic relevance to evaluating the utilitarian value of retrieved evidence, specifically regarding how much a piece of information contributes to reducing the model’s uncertainty during generation \cite{izacard2022atlas}.
Recent studies have highlighted that high semantic similarity between a query and a document does not always guarantee a boost in generation accuracy \cite{infogain}. To address this, researchers have introduced confidence-based utility metrics. For instance, Self-RAG \cite{selfrag} employs critique tokens to evaluate the relevance and supportiveness of retrieved passages, essentially performing a form of implicit information gain assessment. Similarly, FLARE \cite{Flare} and Iter-RetGen \cite{iter-gen} utilize the model’s predictive entropy to decide when to retrieve and which documents provide the most significant knowledge update to the generator’s internal state.
Building upon these confidence-based approaches, InfoGain-RAG \cite{infogain}  introduces Document Information Gain (DIG), a novel metric designed to precisely quantify the contribution of retrieved documents to correct answer generation. DIG measures a document's value by computing the difference in an LLM's generation confidence with and without the document augmented. This framework enables the training of specialized rerankers that prioritize helpful documents while effectively filtering out irrelevant or even misleading content.

% However, document-level metrics like DIG fail to address the internal noise prevalent in long documents or cluttered images, where only a small fragment contains the actual evidence. This coarse granularity inevitably introduces distracting and even misleading information, undermining the generator’s ability to focus on critical information. To overcome this limitation, we extend the information-gain principle to the atomic level and introduce FIG, which quantifies the contribution of individual multimodal units—such as a single sentence or a visual region—thereby maximizing the information density of the retrieved context.
However, document-level metrics such as DIG overlook fine-grained evidence in long documents or cluttered images, introducing noise that distracts the generator. We therefore extend information gain to the atomic level and propose FIG, which measures the contribution of individual multimodal units, such as sentences or visual regions, to maximize the information density of retrieved context.

\section{Task Definition}

We consider the task of MRAG, where a model generates an answer $g$ for a query $x$ by leveraging a large-scale multimodal database $\mathcal{D}$. State-of-the-art MRAG frameworks employ a two-stage retrieve-then-rerank pipeline to select the most relevant context. The end-to-end process can be formally expressed as:
\begin{align}
    \mathcal{D}_{\text{cand}} &= M_{\text{ret}}(x, \mathcal{D}), \label{eq:retrieval} \\
    \mathcal{D}_{\text{sorted}} &= M_{\text{rank}}(x, \mathcal{D}_{\text{cand}}), \label{eq:rerank} \\
    g &= M_{\text{gen}}(x, \mathcal{D}_{\text{sorted}}), \label{eq:generation}
\end{align}
Here, $M_{\text{ret}}$ and $M_{\text{rank}}$ denote the initial retriever and the coarse reranker, respectively, while $M_{\text{gen}}$ represents the downstream MLLM generator.
First, $M_{\text{ret}}$ efficiently recalls a broad candidate set $\mathcal{D}_{\text{cand}}$ from the database. Subsequently, $M_{\text{rank}}$ refines this set by performing a fine-grained relevance assessment, selecting a high-precision subset $\mathcal{D}_{\text{sorted}}$. This refined context is then prepended to the query $x$ and fed into $M_{\text{gen}}$ to produce the final answer $g$.

% \section{Methodology}

% \subsection{Overview: Fine-grained MIG}

% \subsection{Document Segmentation}

% \subsection{Selector}
% defination, goal

% \subsection{Model Distillation}

%（1）介绍 FES-RAG 整体架构流程：描述从传统的“检索-重排”演进为“检索-重排-选择-生成”的四阶段架构 。 （2）介绍文档原子化分解策略：阐述将粗粒度文档拆解为句子和视觉感兴趣区域（RoIs）的必要性 。 （3）介绍递归二分文本分割：详述利用交叉编码器评分进行动态、非固定长度的文本片段提取算法 。 （4）介绍基于 Grounding DINO 的视觉分割：描述如何通过语言引导精准定位并裁剪图像中的关键证据 。 （5）介绍片段信息增益（FIG）的计算：定义并量化单个原子片段对模型生成答案置信度的边际提升 。 （6）介绍 Selector 的训练目标与损失函数：描述二分类选择器及其使用的二元交叉熵（BCE）损失函数 。 （7）介绍知识蒸馏优化机制：阐述如何通过 Teacher-Student 架构将大模型的推理能力迁移至轻量化选择器 。 （8）总结从粗到精的完整推理流水线：整合各模块，描述从海量语库到高密度证据再到纯净生成的全过程 。

\section{Methodology}

\begin{figure*}[ht]
    \centering
    \includegraphics[width=0.95\linewidth]{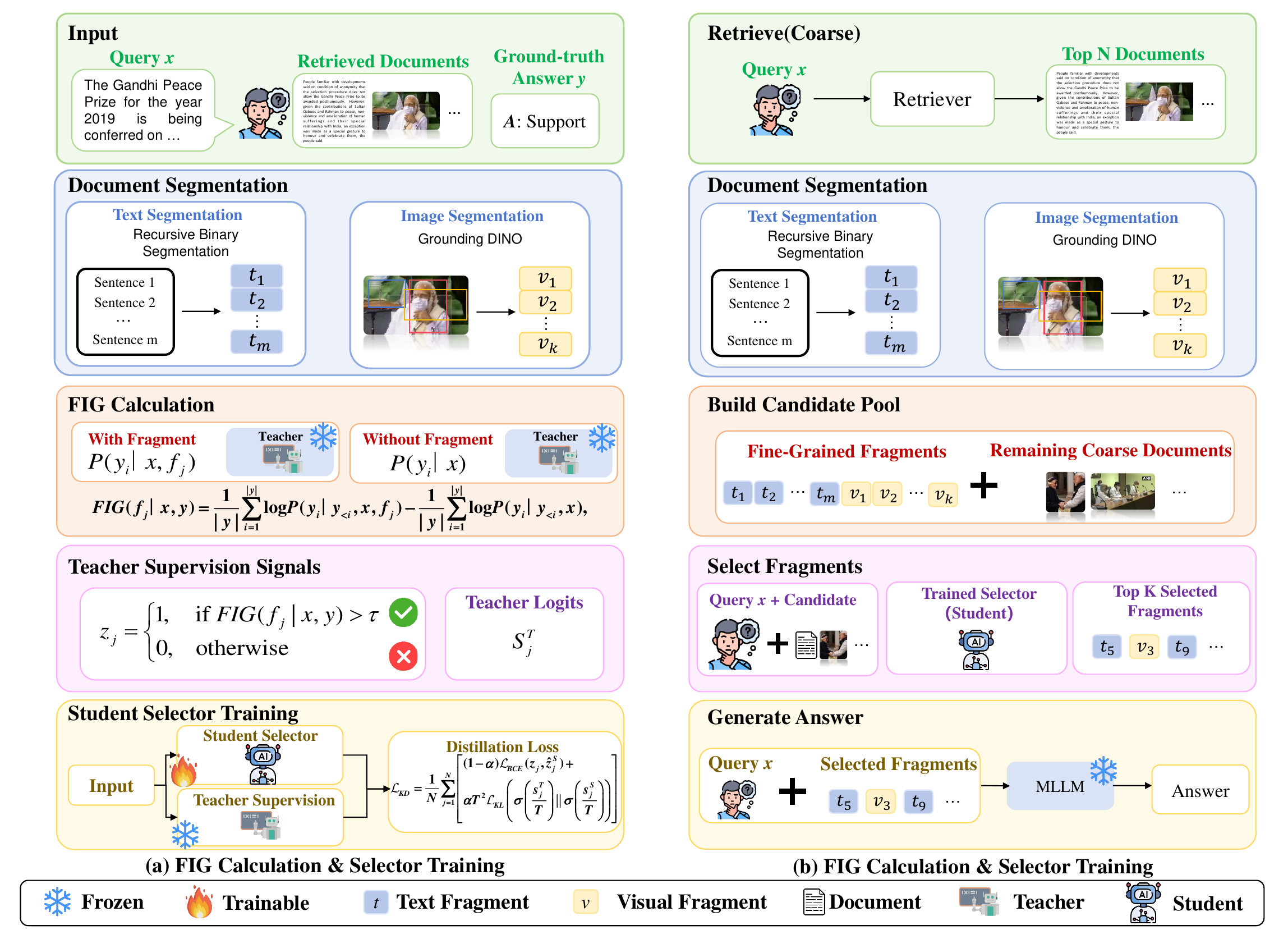}
    % \caption{Overview of the proposed FES-RAG framework. 
    % \textbf{(a) FIG Calculation:} Retrieved documents are first segmented into atomic units---visual ROIs via Grounding DINO and textual sentences via recursive binary segmentation. The FIG is then quantified by measuring the marginal improvement in MLLM confidence provided by each fragment. 
    % \textbf{(b) Selector Training:} A lightweight student selector is optimized via knowledge distillation, simultaneously aligning with the teacher's soft utility distribution via KL divergence and learning from hard labels via binary cross-entropy. 
    % \textbf{(c) FES-RAG Inference:} The pipeline executes a "Retrieve-Segment-Select-Generate" workflow, where the trained selector filters out noise to provide an information-dense, purified context for the final generation.}
  \caption{
Overview of the proposed FES-RAG framework.
\textbf{(a) FIG Calculation and Selector Training:} During training, retrieved multimodal documents are decomposed into textual fragments via recursive binary segmentation and visual fragments via Grounding DINO. A frozen teacher MLLM estimates the Fragment Information Gain (FIG) of each fragment by comparing the likelihood of the ground-truth answer with and without the fragment. The resulting hard labels and teacher logits are then used to train a lightweight student selector through binary classification and knowledge distillation.
\textbf{(b) FES-RAG Inference Pipeline:} During inference, FES-RAG first retrieves top-ranked documents and segments them into fine-grained textual and visual fragments. These fragments are combined with the remaining coarse-grained candidates to form a hybrid evidence pool. The trained student selector scores this pool and selects the top-$K$ high-utility evidence units, which are fed to the MLLM generator as compact, purified context.
}
    \label{fig:method_overview}
\end{figure*}

\subsection{Overview of FES-RAG} 
% Fragment-level Evidence Selection for RAG }
Traditional MRAG systems typically operate at a coarse-grained document level, where entire retrieved documents are provided as context to an MLLM \cite{rora-vlm,re-vilm}. However, full documents often contain significant amounts of redundant information, irrelevant background noise, or contradictory signals that can mislead the generation process \cite{liu2023lost,mei2025survey}.

% 用词与intro没那么统一
To mitigate these issues, we propose the FES-RAG framework, which implements a “Retrieve–Rerank–Select–Generate” pipeline. The introduced selection stage ensures that only informative evidence is presented to the generator. By incorporating a dedicated selector module, FES-RAG performs fine‑grained evidence pruning, extracting only the high‑utility fragments (individual sentences and visual regions) while discarding irrelevant content. This results in a purified, information‑rich context for the final MLLM, thereby improving both factual grounding and computational efficiency.

\subsection{Multimodal Document Segmentation}

%\chengkai{why? previous work infogainrag是怎么做的, lead to shortcomings, -> recall / RAG  ->  }

% Traditional MRAG frameworks primarily treat retrieved evidence as monolithic entities, operating under the implicit assumption that document-level relevance is a sufficient proxy for generative utility \cite{rora-vlm}. However, this coarse-grained approach often encounters a significant \textbf{semantic-utility gap}: a document that is globally relevant to a query may still contain a vast majority of irrelevant background information, filler sentences, or distracting visual artifacts. When these monolithic blocks are prepended to the prompt, they introduce substantial noise that triggers \textbf{attention dilution}, forcing the MLLM to allocate precious computational and cognitive resources to non-informative tokens. This not only increases the risk of the lost-in-the-middle \cite{liu2023lost} phenomenon but also provides fertile ground for hallucinations, as the model may inadvertently synthesize answers from the surrounding distractor context rather than the actual evidence. 

To align the granularity of evidence utilization with the atomic nature of cross-modal reasoning in MLLMs, we propose a document segmentation strategy that breaks down the retrieved document into its most fundamental informative segments. By isolating individual sentences and specific visual ROIs, we can filter out the inherent noise within documents and transform a collection of coarse data into a refined pool of information-dense evidence.

\subsubsection{Textual Segmentation}

To identify the most informative text fragments while filtering out distracting filler content, we propose a Recursive Binary Segmentation strategy. Unlike traditional RAG, which relies on fixed-length sliding windows, our approach dynamically determines the optimal granularity of evidence based on semantic relevance scores.
\begin{algorithm}[t]
\caption{Fine-grained Textual Segmentation via Score-Driven Recursive Decomposition}
\label{alg:text_segmentation}
\begin{algorithmic}[1]
\REQUIRE  {$x$: user query, $t$: retrieved text document, $f_{\phi}$: cross-encoder relevance scorer}
\ENSURE   {$t^*$: the optimal atomic text fragment}

\STATE \textbf{function} \textsc{RecurSplit}($x$, $t$)
    \IF{$t$ is a single sentence}
        \RETURN $t$
    \ENDIF
    \STATE $s_{\text{parent}} \gets f_{\phi}(x, t)$
    \STATE $(t_{\text{left}}, t_{\text{right}}) \gets \text{SplitDoc}(t)$  
    \STATE $s_{\text{left}}  \gets f_{\phi}(x, t_{\text{left}})$   // Left  subsegment score
    \STATE $s_{\text{right}} \gets f_{\phi}(x, t_{\text{right}})$   // Right subsegment score
    \IF{$\max(s_{\text{left}}, s_{\text{right}}) > s_{\text{parent}}$}
        \IF{$s_{\text{left}} > s_{\text{right}}$}
            \RETURN $\textsc{RecurSplit}(x, t_{\text{left}})$
        \ELSE
            \RETURN $\textsc{RecurSplit}(x, t_{\text{right}})$
        \ENDIF
    \ELSE
        \RETURN $t$                                     // No gain, keep intact
    \ENDIF
\STATE \textbf{end function}

\RETURN $\textsc{RecurSplit}(x, t)$
\end{algorithmic}
\end{algorithm}
% \begin{algorithm}[t]
% \caption{Fine-grained Textual Segmentation via Score-Driven Recursive Decomposition}
% \label{alg:text_segmentation}
% \begin{algorithmic}[1]
% \REQUIRE  {$x$: user query, $t$: retrieved text document, $f_{\phi}$: cross-encoder relevance scorer}
% \ENSURE   {$\mathcal{F}$: set of atomic text fragments}

% \STATE $\mathcal{F} \gets \emptyset$                          // Init empty fragment set

% \STATE \textbf{function} \textsc{RecurSplit}($x$, $t$)
%     \IF{$t$ is a single sentence}
%         \RETURN $\{t\}$
%     \ENDIF
%     \STATE $s_{\text{parent}} \gets f_{\phi}(x, t)$
%     \STATE $(t_{\text{left}}, t_{\text{right}}) \gets \text{SplitDoc}(t)$  
%     \STATE $s_{\text{left}}  \gets f_{\phi}(x, t_{\text{left}})$   // Left  subsegment score
%     \STATE $s_{\text{right}} \gets f_{\phi}(x, t_{\text{right}})$   // Right  subsegment score
%     \IF{$\max(s_{\text{left}}, s_{\text{right}}) > s_{\text{parent}}$}
%         \IF{$s_{\text{left}} > s_{\text{right}}$}
%             \RETURN $\textsc{RecurSplit}(x, t_{\text{left}})$
%         \ELSE
%             \RETURN $\textsc{RecurSplit}(x, t_{\text{right}})$
%         \ENDIF
%     \ELSE
%         \RETURN $\{t\}$                                     // No gain, keep intact
%     \ENDIF
% \STATE \textbf{end function}

% \STATE $\mathcal{F} \gets \textsc{RecurSplit}(x, t)$      // Start with full document
% \RETURN $\mathcal{F}$
% \end{algorithmic}
% \end{algorithm}

As shown in Algorithm \ref{alg:text_segmentation}, given a query $x$ and a retrieved text document $t$, we utilize a cross-encoder model (e.g., Jina-Reranker-m0 \cite{jina-m0-blog}) as a scoring function $f_{\phi}(x, t)$, which estimates the relevance of a text segment $t$ relative to $x$. The process begins by calculating the base score of the entire document $s_{parent} = f_{\phi}(x, t)$. We then bisect $t$ into two segments, $s_{left}$ and $s_{right}$, at the sentence boundary nearest to the midpoint. 

The segmentation step is considered gainful if either $f_{\phi}(x, t_{left})$ or $f_{\phi}(x, t_{right})$ exceeds $s_{parent}$. If this condition is met, the document is partitioned, and the process is applied recursively to the partitioned segment with the higher score. The recursion terminates when further subdivision no longer yields a higher relevance score. 
This ensures that the selector receives highly informative fragments, preventing the model from being misled by long, partially irrelevant paragraphs.

% sentence -> coherent semantic unit. 

\subsubsection{Visual Segmentation}

To mitigate the noise interference in coarse-grained retrieved images, we implement a visual segmentation process designed to extract query-relevant local regions. We utilize Grounding DINO~\cite{grounding-dino} as the primary segmentation framework to decompose monolithic images into atomic visual fragments.

Grounding DINO is a cross-modal grounding framework that unifies vision-language understanding with fine-grained object granularity. The architecture comprises a dual-encoder utilizing a Swin Transformer \cite{liu2021swin} for visual feature extraction and BERT \cite{devlin2018bert} for textual encoding—coupled with a feature enhancer for deep cross-modality fusion. By mapping both modalities into a shared latent space, the framework performs language-guided region selection through a cross-modality decoder. This enables the model to identify and segment image areas corresponding to arbitrary textual descriptions in a zero-shot manner, providing a robust foundation for fine-grained evidence extraction without being constrained by predefined object categories.

Formally, let $f_{GD}(x, I)$ denote the Grounding DINO, for a retrieved image $I \in \mathcal{D}_{cand}$ and the user query $x$, the process of generating filtered visual fragments is formulated as follows:
\begin{equation}
\mathcal{B} = \{(b_j, s_j^{\text{obj}}, s_j^{\text{sem}})\}_{j=1}^{n} = f_{GD}(x, I),
\label{eq:detection}
\end{equation}
\begin{equation}
\mathcal{B}_{\text{filtered}} = \{b_j \mid (b_j, s_j^{\text{obj}}, s_j^{\text{sem}}) \in \mathcal{B}, \text{subject to } \mathbf{C} \},
\label{eq:filtering}
\end{equation}
\begin{equation}
\mathbf{C} = \{s_j^{\text{obj}} > \tau_{\text{obj}}, s_j^{\text{sem}} > \tau_{\text{sem}}, \text{Area}(b_j) > \tau_{\text{size}}\},
\label{eq:constraints}
\end{equation}
where $\mathcal{B}$ represents the set of $n$ candidate bounding boxes $b_j$. Each box is associated with two distinct confidence measures: an objectness score $s_j^{obj}$ (representing the probability that the region contains a distinct entity) and a semantic alignment score $s_j^{sem}$ (representing the relevance of the region to the query $x$). The set of constraints $\mathbf{C}$ comprises three complementary criteria designed to ensure the quality of the evidence. First, the objectness constraint ($s_j^{obj} > \tau_{obj}$) filters out spurious regions that the model perceives as non-entities or background artifacts. Second, the semantic alignment constraint ($s_j^{sem} > \tau_{sem}$) ensures that the extracted regions are contextually relevant to the specific user query. Finally, the geometric size constraint ($\text{Area}(b_j) > \tau_{size}$) discards fragments whose pixel area is too small to provide recognizable visual information.

 The regions defined by $\mathcal{B}_{filtered}$ are subsequently cropped from the original image $I$. These purified image patches, denoted as atomic visual fragments, effectively discard irrelevant background pixels and non-target objects, thereby maximizing the marginal utility of the visual context during the final generation phase.

\subsection{FIG Quantification and Teacher Adaptation}

To optimize a high-fidelity selector capable of identifying critical evidence, it is imperative to establish a robust supervision signal that quantifies the intrinsic utility of each candidate fragment. While semantic relevance (e.g., cosine similarity) often serves as a proxy for importance \cite{chen2024bge}, it frequently fails to align with the actual requirements of the generation process. Building upon the principles of DIG \cite{infogain}, we propose FIG, a metric designed to measure the marginal contribution of a fragment to the answer generation.

% 0421
% \subsubsection{FIG Definition}

% Formally, we define FIG as the improvement in the log-likelihood of the ground-truth answer $y$ when the MLLM is augmented with fragment $f_j$:
% \begin{equation}
% FIG(f_j \mid x) = \log P(y \mid x, f_j) - \log P(y \mid x)
% \end{equation}
% where $P(y \mid x, f_j)$ represents the probability of generating the correct response $y$ given the query $x$ and the specific fragment $f_j$, while $P(y \mid x)$ denotes the model’s base confidence derived from its internal parametric knowledge alone. 

% This metric effectively reflects the utility of fragments by filtering out the MLLM's prior biases, thereby providing a clear signal of whether a fragment provides foundational evidence or introduces distracting noise. In our framework, we employ a high-capacity MLLM  \cite{bai2025qwen3vltechnicalreport} to compute these values, leveraging its superior cross-modal reasoning depth to ensure that the resulting utility scores serve as high-fidelity targets for the selector training.

\subsubsection{FIG Definition and Calculation Details}

Formally, we define Fragment Information Gain (FIG) as the improvement in the length-normalized log-likelihood of the ground-truth answer $y$ when the MLLM is augmented with fragment $f_j$:
\begin{equation}
FIG(f_j \mid x, y) = \frac{1}{|y|}\sum_{i=1}^{|y|} \log P(y_i \mid y_{<i}, x, f_j) - \frac{1}{|y|}\sum_{i=1}^{|y|} \log P(y_i \mid y_{<i}, x),
\end{equation}
where $y_i$ is the $i$-th token of the ground-truth answer of length $|y|$, and $y_{<i}$ represents the preceding generated tokens. Here, $P(y_i \mid y_{<i}, x, f_j)$ represents the probability of generating the correct token $y_i$ given the multimodal query $x$, the specific fragment $f_j$, and the previous context, while $P(y_i \mid y_{<i}, x)$ denotes the model’s base confidence derived from its internal parametric knowledge alone. This length normalization is essential to prevent the metric from being disproportionately biased towards shorter responses.

This metric effectively reflects the utility of fragments by filtering out the MLLM's prior biases, thereby providing a clear signal of whether a fragment provides foundational evidence or introduces distracting noise. Furthermore, to capture the model's intrinsic generative confidence without external prompt bias, we compute these probabilities using a zero-shot, open-ended generation format rather than a multi-reference few-shot setup. 

In our framework, we employ a high-capacity MLLM \cite{bai2025qwen3vltechnicalreport} to compute these values, leveraging its superior cross-modal reasoning depth to ensure that the resulting utility scores serve as high-fidelity targets for the selector training.

\subsubsection{Selector Training via Hard Label Supervision}

MLLMs inherently function as generative models rather than discriminators, making them unsuitable for direct fragment-level evidence selection. To address this gap, we first construct high-quality supervision signals and then train a multimodal selector to identify critical evidence fragments without relying on ground-truth answers during inference.

We first compute FIG scores using the methodology described above to quantify the marginal contribution of each fragment $f_j$ to the generation confidence of the ground-truth answer $y$, denoted as $FIG(f_j | x, y)$. These continuous FIG scores are then converted into discrete hard labels $z_j$ using a threshold $\tau$, where $z_j$ denotes whether a fragment is critical for reasoning:
\begin{equation}
z_j = 
\begin{cases} 
1, & \text{if } FIG(f_j | x, y) > \tau \\
0, & \text{otherwise}
\end{cases},
\end{equation}
A label $z_j=1$ indicates that the fragment $f_j$ provides meaningful evidence for answering query $x$, while $z_j=0$ marks the fragment as noise or redundant content. This process yields a supervised dataset where each sample is a tuple $(x, f_j, z_j)$—consisting of a query, a multimodal fragment (text sentence or visual ROIs), and its corresponding binary utility label.

After constructing this supervised dataset, we train the selector on this dataset to map the input pair $(x, f_j)$ to the hard label $z_j$. The model takes only the query $x$ and fragment $f_j$ as input (no ground-truth answer is provided), learning to distinguish between informative fragments and noise purely from cross-modal context. The optimization objective is the Binary Cross-Entropy (BCE) loss:
\begin{equation}
\mathcal{L}_{selector} = -\frac{1}{N} \sum_{j=1}^{N} \left[ z_j \log(\hat{z}_j) + (1 - z_j) \log(1 - \hat{z}_j) \right],
\end{equation}
where $N$ is the total number of fragments in the training set, and $\hat{z}_j = \sigma(f_{\theta}(x, f_j))$ (with $\sigma$ as the sigmoid function) is the selector’s predicted probability that fragment $f_j$ can contribute to the answer generation. 

Unlike rerankers that optimize semantic relevance, minimizing this loss enables the selector to approximate the generator’s internal belief update, where it can reliably identify evidence that directly supports accurate reasoning.

\subsection{Student Selector Optimization via Knowledge Distillation}

While the fine-tuned MLLM Selector demonstrates strong reasoning capability in identifying high-utility atomic fragments, its large parameter count incurs prohibitive inference latency—especially when scoring dozens of atomic fragments (sentences and visual ROIs) per query \cite{flexselect}. 
To address the need for efficient inference in real-world MRAG systems, we propose a knowledge distillation strategy \cite{liu2021semi,liu2021improved,liu2024boosting} to transfer the fragment utility discrimination capability of a high-capacity teacher selector to a lightweight student selector. We employ a lightweight student selector (initialized with Jina-Reranker-2B \cite{jina-m0-blog}) and optimize it using a multi-task knowledge distillation paradigm. This framework allows the student to inherit the precise discrimination capability of the teacher without incurring its computational cost, ensuring both high selection accuracy and efficient inference.

% Specifically, for each atomic fragment $p_j$, the Teacher model (e.g., Qwen3-VL-32B) generates a nuanced utility signal derived from its internal confidence. The Student Selector $f_{\theta}$ is then optimized via a hybrid objective function that enforces probabilistic alignment with the Teacher’s output while minimizing empirical risk. The total distillation loss $\mathcal{L}_{KD}$ is defined as:
The student selector $f_{\theta}$ is trained using a hybrid objective that aligns its predictions with both the ground-truth hard labels and the teacher's soft probability distribution. The total distillation loss $\mathcal{L}_{KD}$ is defined as:
\begin{equation}
\label{eq:kd_loss}
\mathcal{L}_{KD} = \frac{1}{N} \sum_{j=1}^{N} \left[ (1-\alpha) \mathcal{L}_{BCE}(z_j, \hat{z}_j^S) + \alpha T^2 \mathcal{L}_{KL}\left(\sigma\left(\frac{s_j^T}{T}\right) \parallel \sigma\left(\frac{s_j^S}{T}\right)\right) \right],
\end{equation}
where $N$ is the total number of fragments in the training set, $s_j^T$ and $s_j^S$ denote the raw output logits of the teacher and the student model, respectively, while $\hat{z}_j^S = \sigma(s_j^S)$ represents the student’s final predicted probability. The hyperparameter $\alpha$ balances two complementary supervision signals, and $T$ denotes the distillation temperature. 
Specifically, the term $\mathcal{L}_{\text{BCE}}$  employs the binary hard label $z_j$ to enforce hard-label supervision. By providing explicit decision supervision, it guides the student toward the core objective of determining whether a fragment should be selected or discarded, thereby promoting high classification fidelity. However, relying solely on discrete labels can lead to information collapse, where fragments with substantially different utility scores are mapped to the same binary label, and can introduce quantization noise around the threshold $\tau$.
To mitigate these limitations, the term $\mathcal{L}_{\text{KL}}$ enables soft distribution-level distillation by distilling the teacher’s calibrated probability distribution. Regulated by the temperature parameter $T$, these soft targets preserve fine-grained relative utility information, allowing the student to capture nuanced evidence rankings and providing dense, continuous gradient signals even when the predicted class is already correct.

\subsection{FES-RAG Inference}

The inference process of FES-RAG follows a coarse-to-fine trajectory designed to maximize information density while minimizing context noise. The complete procedure is detailed in Algorithm~\ref{alg:mig_rag_distill} and consists of the following core phases:

\begin{itemize}[leftmargin=*]
    \item \textbf{Coarse-grained Retrieval \& Reranking:} Given a query $x$, the system first retrieves $N_{\text{ret}}$ candidates using a retriever $M_{\text{ret}}$ and reorders them via a coarse reranker $M_{\text{rank}}$. This stage effectively narrows the search space from the entire corpus to a manageable set of documents.
    
    \item \textbf{Document Segmentation \& Selection:} To eliminate intra-document noise, the top-ranked items are decomposed into atomic fragments (textual sentences and visual ROIs). To prevent text redundancy, textual documents are entirely replaced by their fragments, whereas original images are retained alongside their visual ROIs to preserve potential global context. Crucially, these fine-grained fragments are combined with the remaining coarse-grained candidates to form a hybrid candidate pool. The distilled selector $M_{\text{distill}}$ then evaluates this unified set, identifying the evidence with the highest predicted FIG.
    
    \item \textbf{Purified Generation:} Instead of feeding entire monolithic documents, only the top-$k$ purified fragments are prepended to the query. The generator $M_{\text{gen}}$ then produces the final answer $g$ based on this information-dense context:
    % \begin{equation}
    %      g = M_{\text{gen}}(q, \text{Select}(\text{Decompose}(D))).
    % \end{equation}
\end{itemize}
By filtering out irrelevant background objects and filler text at the fragment level, this pipeline ensures the MLLM focuses on essential evidence, effectively mitigating the lost-in-the-middle phenomenon while reducing computational overhead.

\begin{algorithm}[t]
\caption{Inference pipeline of FES-RAG}
\label{alg:mig_rag_distill}
\begin{algorithmic}[1]
\REQUIRE $x$: user query, $\mathcal{D}$: multimodal database, $M_{\text{ret}}$: retriever, $M_{\text{rank}}$: coarse reranker, $M_{\text{distill}}$: distilled selector, $M_{\text{gen}}$: generator,
$N_{\text{ret}}$: number of retrieved documents, $N_{\text{seg}}$: number of documents for segmentation, $k$: number of selected fragments for generation
\ENSURE $g$: generated answer

\STATE \textbf{Initialize:} $\mathcal{D}_{\text{aug}} \leftarrow \emptyset$

\STATE \COMMENT{\textbf{Phase 1: Coarse-grained Retrieval \& Reranking}}
\STATE $\mathcal{D}_{\text{cand}} \leftarrow M_{\text{ret}}(x, \mathcal{D}, N_{\text{ret}})$ 
\STATE $\mathcal{D}_{\text{sorted}} \leftarrow M_{\text{rank}}(x, \mathcal{D}_{\text{cand}})$

\STATE \COMMENT{\textbf{Phase 2: Document Segmentation}}
% 2. 明确对前 15 个进行处理
\STATE $\mathcal{D}_{\text{top}} \leftarrow \mathcal{D}_{\text{sorted}}[1 : N_{\text{seg}}]$ 

\FOR{$d \in \mathcal{D}_{\text{top}}$}
    \IF{$\text{IsImage}(d)$}
        \STATE $V_{\text{frags}} \leftarrow \text{ImageSegment}(d)$
        \STATE $\mathcal{D}_{\text{aug}} \leftarrow \mathcal{D}_{\text{aug}} \cup \{d\} \cup V_{\text{frags}}$ 
    \ELSIF{$\text{IsText}(d)$}
        \STATE $T_{\text{frags}} \leftarrow \textsc{RecurSplit}(x, d)$ 
        \STATE $\mathcal{D}_{\text{aug}} \leftarrow \mathcal{D}_{\text{aug}} \cup T_{\text{frags}}$ 
    \ENDIF
\ENDFOR

% 5. 混合细粒度片段和剩余粗粒度文档
\STATE \COMMENT{\textbf{Phase 3: Fine-grained Selection}}
\STATE $\mathcal{D}_{\text{aug}} \leftarrow \mathcal{D}_{\text{aug}} \cup \mathcal{D}_{\text{sorted}}[N_{\text{seg}}+1 : N_{\text{ret}}]$
\STATE $S_{\text{distill}} \leftarrow M_{\text{distill}}(x, \mathcal{D}_{\text{aug}})$  \hfill // Score the utility of the hybrid pool
\STATE $\mathcal{D}_{\text{final}} \leftarrow \text{Sort}(\mathcal{D}_{\text{aug}}, S_{\text{distill}})$

\STATE \COMMENT{\textbf{Phase 4: Purified Generation}}
% 6. 最后只选前 5 个给模型
\STATE $\mathcal{D}_{\text{ctx}} \leftarrow \mathcal{D}_{\text{final}}[1 : k]$ 
\STATE $g \leftarrow M_{\text{gen}}(x, \mathcal{D}_{\text{ctx}})$

\RETURN $g$
\end{algorithmic}
\end{algorithm}
% \vspace{-0.5em}

%（1）介绍实验设置与评价指标：阐述使用的 M²RAG 基准数据集，以及 BLEU、ROUGE 和 CIDEr 等多维度评价指标 。 （2）介绍对比基准模型：列举 None-MRAG、Naïve-MRAG、Jina-Reranker 及粗粒度 InfoGain-MM 等对比基线 。 （3）分析整体 SOTA 性能表现 (RQ1)：展示 FES-RAG 在各项任务中取得的决定性优势，特别是在 CIDEr 分数上的显著提升 。 （4）分析知识蒸馏的增益效果 (RQ2)：对比有无蒸馏的性能差异，证明 Teacher 模型提供的“软标签”比二值标签更有效 。 （5）分析 Top-k 检索深度的鲁棒性：讨论不同片段选择数量对结果的影响，证明细粒度片段具有更强的加性价值 。 （6）分析不同模态下的性能差异 (RQ3)：通过图像子集和文本子集的性能拆解，揭示“视觉瓶颈”并验证片段分割在各模态下的普适性 。 （7）分析效率与 Token 消耗的权衡 (RQ4)：通过可视化实验证明细粒度选择能在降低 Token 消耗的同时提升性能，实现“高密度”生成 。 （8）总结实验关键发现：归纳核心结论，验证细粒度范式在解决注意力稀释和提升事实性方面的核心作用 。

\section{Experiments}

In this section, we conduct comprehensive experiments to evaluate the effectiveness of our enhanced framework, which integrates fine-grained segmentation (visual and textual segmentation) and knowledge distillation. We aim to answer the following research questions:

% 0421
% \begin{itemize}
%     \item \textbf{(RQ1) Overall Performance:} Does our FES-RAG framework outperform existing baselines, including zero-shot, standard RAG, and coarse-grained rerankers across various multimodal tasks?
%     \item \textbf{(RQ2) Ablation Study:} How does the knowledge distillation strategy impact performance, and how sensitive is the model to the number of retrieved segments ($k$)?
%     \item \textbf{(RQ3) Modality Analysis:} How does the system perform when retrieving visual versus textual evidence, and what does this reveal about the retrieval bottleneck?
%     \item \textbf{(RQ4) Efficiency:} Does the fine-grained approach offer a better trade-off between performance and token consumption compared to standard document retrieval?
% \end{itemize}
\begin{itemize}
    \item \textbf{(RQ1) Overall Performance:} Does the proposed FES-RAG consistently outperform existing document-level baselines (e.g., zero-shot, naive retrieval, and coarse-grained reranking) across diverse multimodal tasks and various MLLM architectures?
    % \item \textbf{(RQ2) Ablation \& Validation:} How do the core components—specifically knowledge distillation and candidate depth ($k$)—contribute to the overall performance, and does the proposed Fragment Information Gain (FIG) strictly correlate with the final generation correctness?
    \item \textbf{(RQ2) Ablation \& Validation:} How do the core components, particularly knowledge distillation and candidate depth ($k$), contribute to the overall performance, and how well does the proposed Fragment Information Gain (FIG) align with final generation correctness?
    \item \textbf{(RQ3) Modality Analysis:} How does fine-grained purification impact performance across different modalities (visual vs. textual), and does it inherently resolve the MLLMs' susceptibility to noise better than brute-force truncation?
    \item \textbf{(RQ4) System Efficiency:} How does FES-RAG balance the trade-off between reasoning performance, space efficiency (context token consumption), and time efficiency (end-to-end inference latency) compared to coarse-grained pipelines?
\end{itemize}

\begin{table*}[t!]
\small
\setlength{\tabcolsep}{3pt}
\centering
\caption{Main comparison of FES-RAG-top5 against baselines across multiple MLLMs. \textbf{FES-RAG} utilizes fine-grained segmentation and distillation. The results show that FES-RAG consistently achieves state-of-the-art performance. Best results are in \textbf{bold}, second-best are \underline{underlined}.}
\label{tab:main_results}
\vspace{-1em}
\resizebox{\textwidth}{!}{
\begin{tabular}{ll|cccccccc|cccccccc|cc}
\toprule
\multirow{2}{*}{\textbf{Model}} & \multirow{2}{*}{\textbf{Method}} & \multicolumn{8}{c|}{\textbf{Multimodal QA (MMQA)}} & \multicolumn{8}{c|}{\textbf{Image Captioning}} & \multicolumn{2}{c}{\textbf{Fact Verification}} \\
& & \textbf{B-1} & \textbf{B-2} & \textbf{B-3} & \textbf{B-4} & \textbf{R-1} & \textbf{R-2} & \textbf{R-L} & \textbf{CIDEr} & \textbf{B-1} & \textbf{B-2} & \textbf{B-3} & \textbf{B-4} & \textbf{R-1} & \textbf{R-2} & \textbf{R-L} & \textbf{CIDEr} & \textbf{ACC} & \textbf{F1} \\
\midrule

\multirow{5}{*}{\makecell{MM-RAIT-\\Qwen2.5VL}} 
& None-MRAG & 33.88 & 27.39 & 23.15 & 20.20 & 48.31 & 30.52 & 46.25 & 206.42 & 16.18 & 10.46 & 7.56 & 5.68 & 34.16 & 15.43 & 32.58 & 79.41 & 59 & 61.15 \\
& Naïve-MRAG-top3 & 40.67 & 31.39 & 26.08 & 23.07 & 56.56 & 33.37 & 54.09 & 247.13 & 17.15 & 11.41 & 7.90 & 5.89 & 36.09 & 16.89 & 34.57 & 84.41 & 62 & 62.95 \\
& Jina-Reranker-top3 & 40.46 & 31.61 & 26.29 & 23.41 & 56.34 & 34.17 & 54.05 & 248.63 & 19.39 & 13.05 & 9.47 & 7.42 & 38.05 & 18.31 & 35.74 & 96.66 & 63 & 64.31 \\
& InfoGain-MM-top3 & \underline{42.95} & \underline{33.38} & \underline{27.60} & \underline{24.14} & \underline{61.24} & \underline{35.89} & \underline{57.91} & \underline{264.78} & \underline{23.47} & \underline{17.46} & \underline{13.74} & \underline{11.36} & \underline{42.91} & \underline{24.10} & \underline{41.24} & \underline{136.44} & \underline{68} & \underline{69.80} \\
& \textbf{FES-RAG-top5} & \textbf{49.31} & \textbf{39.48} & \textbf{34.60} & \textbf{31.35} & \textbf{67.94} & \textbf{42.95} & \textbf{65.93} & \textbf{336.42} & \textbf{27.83} & \textbf{21.44} & \textbf{17.58} & \textbf{14.96} & \textbf{48.13} & \textbf{29.35} & \textbf{46.57} & \textbf{171.51} & \textbf{70} & \textbf{72.71} \\
\midrule

\multirow{5}{*}{\makecell{LLaVA-NeXT}} 
& None-MRAG & 34.85 & 30.00 & 27.04 & 24.49 & 48.36 & 37.00 & 45.61 & 250.82 & 6.42 & 3.08 & 2.02 & 1.63 & 19.17 & 5.34 & 17.33 & 18.98 & 39 & 35.76 \\
& Naïve-MRAG-top3 & 35.73 & 30.81 & 27.65 & 24.90 & 49.81 & 37.83 & 46.93 & 250.52 & 9.30 & 5.52 & 3.85 & 2.84 & 23.99 & 9.74 & 21.87 & 35.48 & 47 & 31.14 \\
& Jina-Reranker-top3 & 36.19 & 31.37 & 28.32 & 25.78 & 50.08 & 38.37 & 47.44 & 259.41 & 10.05 & 5.98 & 4.10 & 3.02 & 25.79 & 10.68 & 22.91 & 44.60 & 50 & 34.26 \\
& InfoGain-MM-top3 & \underline{37.89} & \underline{33.16} & \underline{30.21} & \underline{27.71} & \underline{52.66} & \underline{41.01} & \underline{50.22} & \underline{278.43} & \underline{13.17} & \underline{8.18} & \underline{5.95} & \underline{4.43} & \underline{29.85} & \underline{12.84} & \underline{25.96} & \underline{57.78} & \underline{55} & \underline{38.25} \\
& \textbf{FES-RAG-top5} & \textbf{43.26} & \textbf{35.12} & \textbf{31.18} & \textbf{28.42} & \textbf{60.72} & \textbf{44.50} & \textbf{57.79} & \textbf{299.29} & \textbf{16.16} & \textbf{11.26} & \textbf{8.33} & \textbf{6.87} & \textbf{36.23} & \textbf{18.53} & \textbf{34.01} & \textbf{92.82} & \textbf{65} & \textbf{65.69} \\
\midrule

\multirow{5}{*}{\makecell{InternVL3.5}} 
& None-MRAG & 25.34 & 19.78 & 16.75 & 14.44 & 38.82 & 26.67 & 34.92 & 114.24 & 11.76 & 6.69 & 4.37 & 3.07 & 23.66 & 9.42 & 21.47 & 45.09 & 41 & 38.96 \\
& Naïve-MRAG-top3 & 29.57 & 24.12 & 20.32 & 17.84 & 46.52 & 31.99 & 43.66 & 156.10 & 10.96 & 6.24 & 3.61 & 2.75 & 27.12 & 12.43 & 24.33 & 41.46 & 47 & 46.89 \\
& Jina-Reranker-top3 & 30.57 & 25.17 & 21.29 & 18.53 & 47.66 & 33.21 & 44.42 & 166.41 & 11.43 & 6.63 & 3.94 & 2.99 & 27.91 & 11.98 & 24.11 & 40.73 & 52 & 52.02 \\
& InfoGain-MM-top3 & \underline{34.26} & \underline{28.42} & \underline{25.04} & \underline{22.25} & \underline{51.44} & \underline{36.57} & \underline{49.33} & \underline{219.21} & \underline{15.12} & \underline{8.64} & \underline{5.84} & \underline{4.79} & \underline{32.04} & \underline{15.38} & \underline{28.29} & \underline{64.00} & \underline{55} & \underline{55.04} \\
& \textbf{FES-RAG-top5} & \textbf{41.77} & \textbf{33.19} & \textbf{28.27} & \textbf{25.48} & \textbf{61.58} & \textbf{37.34} & \textbf{59.05} & \textbf{264.87} & \textbf{18.60} & \textbf{13.26} & \textbf{9.94} & \textbf{8.07} & \textbf{39.09} & \textbf{20.40} & \textbf{37.18} & \textbf{108.18} & \textbf{67} & \textbf{68.62} \\
\midrule

\multirow{5}{*}{\makecell{Qwen2.5-VL}} 
& None-MRAG & 25.41 & 20.64 & 17.61 & 14.91 & 39.30 & 27.78 & 34.97 & 105.33 & 8.97 & 4.46 & 2.84 & 2.03 & 20.05 & 6.64 & 18.20 & 19.51 & 51 & 50.67 \\
& Naïve-MRAG-top3 & 30.67 & 25.45 & 22.14 & 19.55 & 45.04 & 32.63 & 41.42 & 185.94 & 11.75 & 6.92 & 4.93 & 3.70 & 26.99 & 10.77 & 24.55 & 46.64 & 60 & 62.26 \\
& Jina-Reranker-top3 & 29.89 & 25.19 & 22.16 & 19.68 & 45.41 & 33.50 & 41.39 & 174.55 & 12.62 & 7.16 & 4.56 & 3.49 & 27.82 & 11.86 & 25.43 & 45.73 & 63 & 65.96 \\
& InfoGain-MM-top3 & \underline{33.37} & \underline{28.11} & \underline{24.79} & \underline{22.02} & \underline{47.94} & \underline{36.03} & \underline{44.73} & \underline{218.62} & \underline{15.31} & \underline{9.77} & \underline{7.10} & \underline{5.58} & \underline{31.24} & \underline{16.29} & \underline{28.50} & \underline{73.66} & \underline{70} & \underline{73.75} \\
& \textbf{FES-RAG-top5} & \textbf{40.29} & \textbf{32.11} & \textbf{27.38} & \textbf{24.73} & \textbf{60.43} & \textbf{40.23} & \textbf{58.18} & \textbf{256.36} & \textbf{20.51} & \textbf{14.77} & \textbf{11.30} & \textbf{8.99} & \textbf{39.89} & \textbf{21.23} & \textbf{37.02} & \textbf{110.99} & \textbf{73} & \textbf{74.97} \\
\bottomrule
\end{tabular}
}
\end{table*}
% =================================================================================
% Experimental Setup
% =================================================================================
\subsection{Experimental Setup}
% \noindent
 \textbf{Datasets.} We evaluate our framework on the \textbf{M2RAG} benchmark~\cite{liu2025benchmarking}, a challenging large-scale dataset designed to assess retrieval-augmented generation capabilities across heterogeneous modalities. The benchmark simulates realistic open-domain scenarios by including a mix of relevant, irrelevant, and partially relevant multimodal documents. It encompasses three distinct tasks: (1) \textbf{Multi-Modal QA}, requiring models to synthesize answers from visual and textual evidence; (2) \textbf{Image Captioning}, which tests the ability to generate descriptive text conditioned on retrieved visual context; and (3) \textbf{Fact Verification}, a classification task determining whether multimodal evidence supports a given claim.

\textbf{Evaluation Metrics.} We employ comprehensive metrics to evaluate generation quality and reasoning accuracy. For generative tasks (MMQA and Image Captioning), we report BLEU-1/2/3/4~\cite{papineni2002bleu} and ROUGE-1/2/L~\cite{lin2004rouge}, which quantify n-gram precision and recall against ground-truth references. Additionally, we utilize CIDEr~\cite{vedantam2015cider}, a metric specifically designed for image captioning that captures consensus-based semantic similarity and has been shown to correlate better with human judgment in vision-language tasks. For the fact verification task, performance is assessed using standard Accuracy and F1 Score to evaluate classification precision.

\textbf{Baselines.} To verify the universality of our approach, we first employ diverse backbone MLLMs, including open-source MLLMs (\textbf{Qwen2.5-VL}\cite{bai2025qwen2}, \textbf{LLaVA-NeXT}\cite{li2024llava}, \textbf{InternVL3.5}\cite{wang2025internvl3}) as well as \textbf{MM-RAIT-Qwen2.5VL}\cite{liu2025benchmarking}, a specialized baseline fine-tuned on the M2RAG benchmark for retrieval-augmented tasks. 
Based on these generators, we compare our proposed method against four retrieval baselines: (1) \textit{None-MRAG} (zero-shot inference without retrieval); (2) \textit{Naïve-MRAG} (Top-3 retrieval using raw embeddings); (3) \textit{Jina-Reranker} (off-the-shelf Top-3 without fine-tuning); and (4) \textit{InfoGain-MM}, which extends the InfoGain-RAG \cite{infogain} to the multimodal domain (serving as our coarse-grained Top-3 baseline). Our proposed method, denoted as FES-RAG, utilizes fine-grained segmentation and distillation. Note that we compare the baseline's Top-3 documents against our Top-5 segments, as they incur comparable token consumption, ensuring a fair evaluation under a strictly controlled context window budget.

\textbf{Implementation Details.} 
% Experiments were conducted on 8 NVIDIA A100 GPUs with 80GB of memory each. The training dataset was approximately balanced, containing 10k positive samples and 12k negative samples. For the initial retrieval stage ($M_{\text{ret}}$), we employed Visualized BGE~\cite{zhou2024vista}, a state-of-the-art multimodal dense retriever, to encode both queries and documents into a shared embedding space. We implemented the knowledge distillation strategy where a Qwen3-VL-32B~\cite{bai2025qwen3vltechnicalreport} teacher guides the Jina-Reranker-m0~\cite{jina-m0-blog} student. We selected this model due to its widespread adoption within the research community and its proven effectiveness across various multimodal benchmarks. The student model was optimized to minimize Eq. \ref{eq:kd_loss} with distillation weight $\alpha=0.7$ and temperature $T=2$. The training was performed for 5 epochs with a global batch size of 32, utilizing an AdamW optimizer \cite{loshchilov2017decoupled} with a learning rate of $2e^{-5}$.
% Regarding the inference hyperparameters, we set the initial retrieval depth $N_{\text{ret}} = 100$ to ensure high recall. The segmentation budget is set to $N_{\text{seg}} = 15$, prioritizing the top-ranked candidates for document segmentation to balance efficiency and coverage.
% For the selector training, the hard label generation threshold $\tau$ is set to 0.2. 
% For grounding DINO segmentation, we adopt filtering thresholds of
% $\tau_{\mathrm{obj}}=0.40$, $\tau_{\mathrm{sem}}=0.35$, 
% and $\tau_{\mathrm{size}}=50^{2}~\mathrm{px}^{2}$ for the bounding-box area.
Experiments are conducted on 8 NVIDIA A100 GPUs with 80GB of memory each. The training dataset is approximately balanced, containing 10k positive samples and 12k negative samples. For the initial retrieval stage ($M_{\text{ret}}$), we employ Visualized BGE~\cite{zhou2024vista}, a state-of-the-art multimodal dense retriever, to encode both queries and documents into a shared embedding space. We implement the knowledge distillation strategy where a Qwen3-VL-32B~\cite{bai2025qwen3vltechnicalreport} teacher guides the Jina-Reranker-m0~\cite{jina-m0-blog} student. We select Qwen3-VL-32B as the teacher due to its strong cross-modal reasoning capability and proven effectiveness across various multimodal benchmarks. The student model is optimized to minimize Eq.~\ref{eq:kd_loss} with distillation weight $\alpha=0.7$ and temperature $T=2$. Training is performed for 5 epochs with a global batch size of 32, using the AdamW optimizer~\cite{loshchilov2017decoupled} with a learning rate of $2e^{-5}$.
Regarding the inference hyperparameters, we set the initial retrieval depth to $N_{\text{ret}}=100$ to ensure high recall. The segmentation budget is set to $N_{\text{seg}}=15$, prioritizing the top-ranked candidates for document segmentation to balance efficiency and coverage. For selector training, the hard-label generation threshold $\tau$ is set to 0.2. For Grounding DINO segmentation, we adopt filtering thresholds of $\tau_{\mathrm{obj}}=0.40$, $\tau_{\mathrm{sem}}=0.35$, and $\tau_{\mathrm{size}}=2{,}500~\mathrm{px}^{2}$
\subsection{Overall Performance (RQ1)}
Table \ref{tab:main_results} reports the comparative performance across three distinct multimodal tasks and four MLLMs. The results reveal several critical insights into the limitations of current MRAG systems and the efficacy of FES-RAG: \textbf{(i) The Failure of Generic Reranking.} While \textit{Naïve-MRAG} significantly boosts performance over the zero-shot baseline (\textit{None-MRAG}), the off-the-shelf \textit{Jina-Reranker} yields marginal or even negligible gains (e.g., only +1.5 CIDEr on MMQA with MM-RAIT). This empirically confirms a misalignment in existing retrieval paradigms: generic rerankers optimize for semantic similarity, which does not necessarily translate to reasoning utility. A document can be semantically similar to a query yet lack the specific visual or textual evidence required to answer it. \textbf{(ii) Granularity Matters.} While our coarse-grained baseline, \textit{InfoGain-MM}, surpasses standard rerankers by optimizing for utility, it remains inferior to FES-RAG. The superiority of fine-grained retrieval is evidenced by the substantial performance gap on the MMQA benchmark, where FES-RAG-top5 increases the CIDEr score from 264.78 to 336.42, representing a notable 27\% relative improvement over the baseline. This empirical evidence validates our core hypothesis that document-level retrieval is insufficient for fine-grained reasoning. By decomposing documents into atomic units, FES-RAG effectively eliminates contextual clutter, such as irrelevant visual backgrounds and filler sentences, thereby preventing the noise interference. \textbf{(iii) Universality Across Architectures.} Crucially, the superiority of FES-RAG is consistent across all tested backend generators, ranging from the specialized MM-RAIT to general-purpose models like LLaVA-NeXT and InternVL3.5. This suggests that susceptibility to contextual noise is an intrinsic characteristic of MLLMs regardless of their architecture size, and providing purified, high-density evidence serves as a universally effective strategy for enhancing factual grounding.

\subsection{Ablation Study (RQ2)}
\subsubsection{Effectiveness of Knowledge Distillation}

\begin{figure}[h!]
    \centering
    \includegraphics[width=\linewidth]{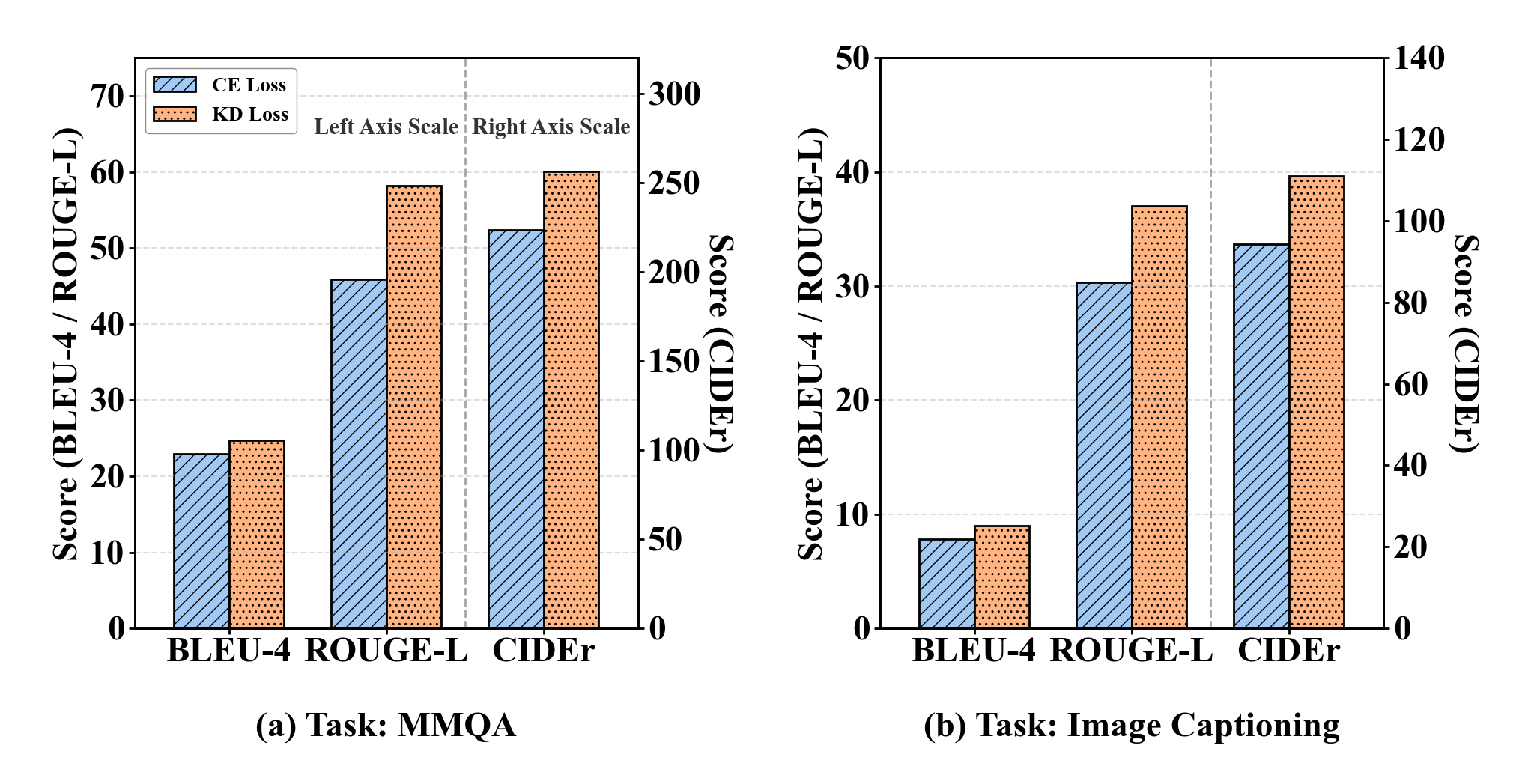}
    \vspace{-2em}
    \caption{\textbf{Impact of Knowledge Distillation.} Comparison among three identical Jina-Reranker-2B student selectors: trained with hard labels (\textbf{CE (No Teacher)}), distilled from a same-size teacher (\textbf{Same-Size KD} using Qwen3-VL-2B), and distilled from the full teacher (\textbf{FES-RAG} using Qwen3-VL-32B). The results confirm the intrinsic value of soft distribution-level distillation independent of model scaling.}
    \label{fig:ablation_distillation}
\end{figure}

To validate the efficacy of our supervision strategy, we compare identical student selectors (Jina-Reranker-2B) trained under different objectives. As illustrated in Figure \ref{fig:ablation_distillation}, models trained solely with standard Cross-Entropy (\textbf{CE (No Teacher)}) rely exclusively on hard-label supervision, which projects continuous utility scores into discrete binary targets. This rigid quantization inevitably causes information collapse, stripping away the fine-grained relative ranking information among candidate fragments. 

To isolate the intrinsic benefits of our distillation formulation from the raw parameter advantage of a massive teacher model, we introduce a same-size KD setup where the student is supervised by a teacher of identical capacity (Qwen3-VL-2B). Even without a parameter gap, the Same-Size KD student (CIDEr 242.60) consistently outperforms the CE-only baseline (CIDEr 223.38). We attribute this performance leap to the synergistic integration of soft distribution-level distillation. By distilling the calibrated soft logits from the teacher, the student captures nuanced utility assessments, enabling it to distinguish subtle high-quality evidence from hard negatives more effectively than binary supervision alone. 

Naturally, scaling the teacher to the full Qwen3-VL-32B model (\textbf{FES-RAG}) further elevates the upper bound of this capability, achieving a peak CIDEr score of 256.36 on the MMQA task.

% To validate the benefit of distilling knowledge from Qwen3-VL, we compared two fine-grained student selectors: CE Loss (trained with standard Cross-Entropy) and KD Loss (trained with our distillation objective). 
% As shown in Figure \ref{fig:ablation_distillation}, the student selector trained with KD Loss consistently outperforms the CE Loss variant across all metrics. Notably, on MMQA, the CIDEr score (Right Axis) improves from 223.38 to \textbf{256.36}. This indicates that the teacher model provides softer, more nuanced supervision signals than binary labels alone, helping the selector better distinguish subtle relevance.
\subsubsection{Robustness to Candidate Selection (Top-k)}
% We analyze the scaling behavior of our method. Table \ref{tab:topk_ablation} compares our method against the coarse-grained baseline at different $k$ levels.
% A striking finding is that FES-RAG-top1 (CIDEr 237.49) already surpasses the InfoGain-MM-top3 baseline (218.62). This implies that our reranker is highly effective at placing the most critical evidence in the first position. Furthermore, performance steadily increases from $k=1$ to $k=5$, showing that our fine-grained segments provide additive value without introducing overwhelming noise.
We further examine the scaling behavior of our method with respect to the number of retrieved segments $k$, as presented in Table \ref{tab:topk_ablation}. The results highlight the exceptional discriminative capability of our distilled student selector, where FES-RAG-top1 alone achieves a CIDEr score of 237.49, notably surpassing the coarse-grained InfoGain-MM-top3 baseline of 218.62. This finding underscores that our selector effectively prioritizes the most critical evidence, enabling superior reasoning capabilities even with minimal context. Moreover, as $k$ increases from 1 to 5, the performance shows a robust upward trajectory. Although expanding the context window in coarse-grained retrieval also yields performance gains by accumulating more potential evidence, it simultaneously introduces substantial noise, limiting the reasoning accuracy. In contrast, FES-RAG mitigates this issue by enforcing high information density at the fragment level. Unlike coarse-grained approaches that rely on larger context windows to capture sufficient evidence, our method ensures that each additional atomic fragment selected by the student selector provides high-utility information, thereby maximizing the marginal gain of the retrieval budget while minimizing noise interference.

\begin{table}[H]
\centering
\small
\setlength{\tabcolsep}{4pt}
\caption{Comparison of Top-k scaling strategies on Qwen2.5-VL. Best results are in bold.}
\label{tab:topk_ablation}
\vspace{-1em}
\resizebox{\columnwidth}{!}{
\begin{tabular}{l|ccc|ccc}
\toprule
\multirow{2}{*}{\textbf{Method}} & \multicolumn{3}{c|}{\textbf{MMQA}} & \multicolumn{3}{c}{\textbf{Image Captioning}} \\
& \textbf{B-4} & \textbf{R-L} & \textbf{CIDEr} & \textbf{B-4} & \textbf{R-L} & \textbf{CIDEr} \\
\midrule
InfoGain-MM-top1 & 21.85 & 44.17 & 211.24 & 4.12 & 26.62 & 50.00 \\
\textbf{FES-RAG-top1} & \textbf{22.25} & \textbf{54.62} & \textbf{237.49} & \textbf{7.13} & \textbf{35.00} & \textbf{96.32} \\
\midrule
InfoGain-MM-top3 & 22.02 & 44.73 & 218.62 & 5.58 & 28.50 & 73.66 \\
\textbf{FES-RAG-top3} & \textbf{23.43} & \textbf{55.29} & \textbf{247.94} & \textbf{8.62} & \textbf{36.16} & \textbf{107.38} \\
\midrule
InfoGain-MM-top5 & 24.57 & 45.80 & 239.79 & 8.56 & 31.87 & 103.80 \\
\textbf{FES-RAG-top5} & \textbf{24.73} & \textbf{58.18} & \textbf{256.36} & \textbf{8.99} & \textbf{37.02} & \textbf{110.99} \\
\bottomrule
\end{tabular}
}
\end{table}

\subsubsection{Correlation between FIG and Final Correctness}
To empirically validate that Fragment Information Gain (FIG) monotonically aligns with downstream correctness—rather than merely serving as a proxy for ground-truth log-likelihood—we conduct a controlled Top-1 injection sanity check. 

To rigorously evaluate the final correctness of open-ended generative responses in the MMQA task, we employ GLM-4.6V \cite{hong2025glm} as an \textbf{LLM-as-a-Judge} evaluator. The evaluator assesses whether the MLLM's generated response semantically entails the ground truth, yielding a semantic Accuracy (Acc) score. This metric effectively captures the true accuracy gain while avoiding the rigid false-negative penalties associated with exact string matching.

\begin{table}[ht]
\centering
\small
\setlength{\tabcolsep}{8pt}
\caption{Sanity check showing the correlation between FIG score intervals and final generative correctness (LLM Judge Accuracy and CIDEr) on the MMQA task (Qwen2.5-VL).}
\label{tab:fig_correctness}
\vspace{-1em}
\begin{tabular}{l|cc}
\toprule
\textbf{FIG Interval} & \textbf{LLM Judge Acc \%} & \textbf{CIDEr} \\
\midrule
$FIG \le 0.0$ & 32 & 102.45 \\
$0.0 < FIG \le 0.2$ & 65 & 182.95 \\
$FIG > 0.2$ & \textbf{92} & \textbf{237.49} \\
\bottomrule
\end{tabular}
\end{table}
We force the Qwen2.5-VL generator to condition on exactly one fragment from a specific FIG interval. As shown in Table \ref{tab:fig_correctness}, the generative accuracy strictly aligns with the FIG intervals. When conditioned on noise fragments ($FIG \le 0.0$), the model's accuracy is restricted to 32\% (CIDEr 102.45), confirming these fragments fail to provide grounding and force the model to rely solely on internal parametric guesses. For fragments with low to moderate utility ($0.0 < FIG \le 0.2$), the accuracy improves to 65\%. Conversely, when conditioned on high-utility fragments ($FIG > 0.2$) successfully isolated by FES-RAG, the LLM Judge Accuracy surges to an impressive 92\%, accompanied by a peak CIDEr score of 237.49. These results provide empirical evidence that higher FIG scores are associated with stronger downstream generation correctness, supporting FIG as an effective supervision signal for fragment-level evidence selection.

\subsection{Modality Analysis (RQ3)}
To deconstruct the source of our performance gains, we partition the test sets into image and text subsets and evaluate the backbone Qwen2.5-VL across multiple tasks, as shown in Table \ref{tab:modality_ablation}.

The results highlight that FES-RAG's purification is effective across all scenarios. On the \textbf{Image Subset}, baseline scores are relatively high, yet FES-RAG still achieves a significant boost, increasing the MMQA CIDEr score from 427.49 to 439.59. Standard MLLMs typically perform image downsampling to fit input resolution constraints, often causing small but critical objects to become unrecognizable. By feeding explicit ROIs, FES-RAG preserves high-resolution visual details, preventing resolution dilution and thereby boosting FactVeri ACC from 72.73\% to 75.76\%.

The impact is even more pronounced on the \textbf{Text Subset}, where FES-RAG causes MMQA CIDEr scores to more than double (from 46.33 to 106.44) alongside a significant improvement in FactVeri F1 score (from 65.10 to 66.41). To verify that this massive gain stems from our precise \textbf{Recursive Binary Segmentation} rather than simply reducing the context window length, we introduce a \textbf{Brute-force Truncation} baseline. This baseline directly truncates the originally retrieved documents to match the exact token budget of our selected fragments, entirely bypassing the FIG-guided selection. As shown in Table \ref{tab:modality_ablation}, while brute-force truncation yields a minor improvement (e.g., CIDEr 62.18) by casually filtering some distal noise, it falls massively short of FES-RAG. This disparity confirms that MLLMs are highly sensitive to textual noise. Whereas coarse-grained or arbitrarily truncated documents obscure the precise factual evidence within paragraphs of lengthy distractors, our fragment-level selection effectively isolates the critical facts, thereby purifying the reasoning context for diverse tasks.

\begin{table}[H]
\centering
\small
\setlength{\tabcolsep}{3pt}
\caption{Performance breakdown by retrieved modality subsets across MMQA and Fact Verification (Qwen2.5-VL).}
\label{tab:modality_ablation}
\resizebox{\columnwidth}{!}{
\begin{tabular}{l|l|ccc|cc}
\toprule
\multirow{2}{*}{\textbf{Subset}} & \multirow{2}{*}{\textbf{Method}} & \multicolumn{3}{c|}{\textbf{MMQA}} & \multicolumn{2}{c}{\textbf{Fact Verification}} \\
& & \textbf{B-4} & \textbf{R-L} & \textbf{CIDEr} & \textbf{ACC \%} & \textbf{F1} \\
\midrule
\multirow{2}{*}{\textbf{Image}} 
& InfoGain-MM-top3 & 42.52 & 73.78 & 427.49 & 72.73 & 78.21 \\
& \textbf{FES-RAG-top5} & \textbf{45.09} & \textbf{74.57} & \textbf{439.59} & \textbf{75.76} & \textbf{79.38} \\
\midrule
\multirow{3}{*}{\textbf{Text}} 
& InfoGain-MM-top3 & 5.24 & 20.95 & 46.33 & 64.71 & 65.10 \\
& Brute-force Truncation & 6.15 & 27.40 & 62.18 & 64.71 & 65.82 \\
& \textbf{FES-RAG-top5} & \textbf{8.07} & \textbf{44.76} & \textbf{106.44} & \textbf{67.65} & \textbf{66.41} \\
\bottomrule
\end{tabular}
}
\end{table}

% =================================================================================
% RQ4: Efficiency
% =================================================================================

% 0421
% \subsection{Efficiency Analysis (RQ4)}

% A potential concern with increasing $k$ is the computational cost. We visualize the trade-off between average token consumption and model performance in Figure \ref{fig:token_efficiency}.
% Contrary to the assumption that higher $k$ implies higher cost, the results reveal a clear efficiency advantage: FES-RAG-top5 consumes fewer tokens (874.29) compared to the baseline InfoGain-MM-top3 (1098.89), represented by the lower green bar, while achieving significantly higher performance (red line).
% This result is pivotal: demonstrating that the performance gains stem from higher information density rather than increased context length. Specifically, the fine-grained segmentation successfully filters out irrelevant noise, resulting in a context that is both richer in information and lighter in computation. This low-token, high-information property confirms the efficiency of FES-RAG.
% \begin{figure}[h!]
%     \centering
%     \includegraphics[width=\linewidth]{Fig/token_efficiency.png}
%     % \vspace{-1em}
%     \caption{\textbf{Efficiency Analysis:} Comparison of token consumption vs. performance. Our fine-grained strategy reduces tokens while boosting accuracy.}
%     \label{fig:token_efficiency}
%     % \vspace{-2em}
% \end{figure}

\subsection{Efficiency Analysis (RQ4)}

A potential concern with introducing a multi-stage fragment-level pipeline is the incremental computational overhead. To address this, we evaluate FES-RAG's efficiency across space (token consumption) and time (inference latency).

\begin{figure}[htbp]
    \centering
    \includegraphics[width=\linewidth]{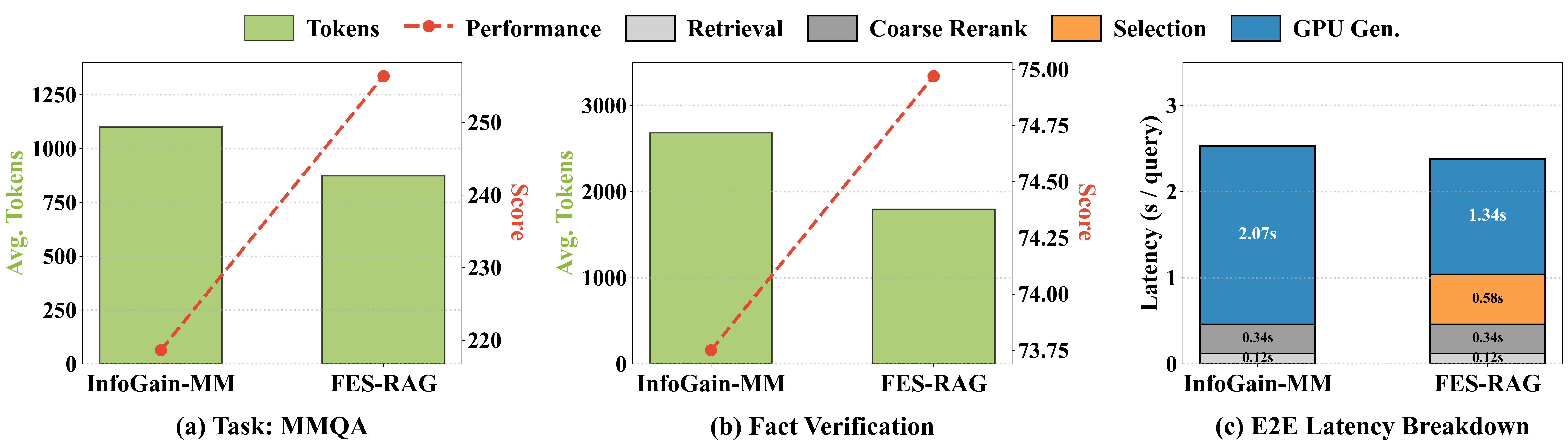}
    \caption{\textbf{Efficiency and Performance Trade-offs.} Comparison of context token consumption (a, b) and end-to-end inference latency (c).}
    \label{fig:efficiency_analysis}
\end{figure}

As illustrated in Figure \ref{fig:efficiency_analysis}, FES-RAG demonstrates a significant dual-efficiency advantage. First, in terms of space efficiency (Fig. \ref{fig:efficiency_analysis}a,b), our framework achieves substantially higher performance while reducing context length by 20.4\% in MMQA and 33.1\% in Fact Verification compared to the coarse-grained InfoGain-MM baseline. This confirms that fragment-level segmentation successfully purifies the context, maximizing information density.

Second, this reduction directly translates into a remarkable inference acceleration (Fig. \ref{fig:efficiency_analysis}c). We decompose the End-to-End (E2E) latency into four distinct phases: Vector Retrieval, Coarse Reranking, Segmentation \& Selection, and GPU Generation. While the coarse-grained InfoGain-MM proceeds directly to generation after reranking, FES-RAG introduces a fine-grained selection stage (0.58s) to filter redundant information. Crucially, this purification drastically lightens the computational load on the heavy MLLM generator. 
Since the self-attention cost during the prefill stage grows quadratically with the input context length, reducing the retrieved context can substantially lower generation latency. By filtering redundant evidence before generation, FES-RAG reduces the generation time from 2.07s to 1.34s. The latency saved during generation (+0.73s) outweighs the additional overhead introduced by the selector (-0.58s), leading to a lower overall end-to-end latency than the coarse-grained baseline (2.38s vs. 2.53s). These results suggest that fragment-level purification can serve as a system-level accelerator, making FES-RAG suitable for latency-sensitive applications.
% Since the attention complexity of MLLMs scales quadratically with sequence length, FES-RAG slashes the generation time from 2.07s down to 1.34s. The latency saved during the generation phase (+0.73s) significantly outweighs the overhead of our selector (-0.58s), resulting in a lower overall E2E latency (2.38s vs. 2.53s). These results suggest that fragment-level purification acts as a system-level accelerator, making FES-RAG highly deployable for latency-sensitive applications.

% --- START OF CASE STUDY SECTION ---

% 定义子模块的样式，让代码更整洁
% \newtcolorbox{casebox}[1]{
%     colback=white,
%     colframe=gray!20,
%     fonttitle=\bfseries\small,
%     coltitle=black,
%     title={#1},
%     boxrule=1pt,
%     sharp corners,
%     rounded corners=southeast,
%     arc=4pt,
%     left=2pt, right=2pt, top=2pt, bottom=2pt
% }

% % 定义用于显示Label的小命令
% \newcommand{\caselabel}[1]{\textbf{\textcolor{blue!60!black}{\small [Label=#1]}} }

\section{Qualitative Results: Case Study}
\label{sec:appendix_cases}

\begin{figure}[t]
    \centering
    \includegraphics[width=\linewidth]{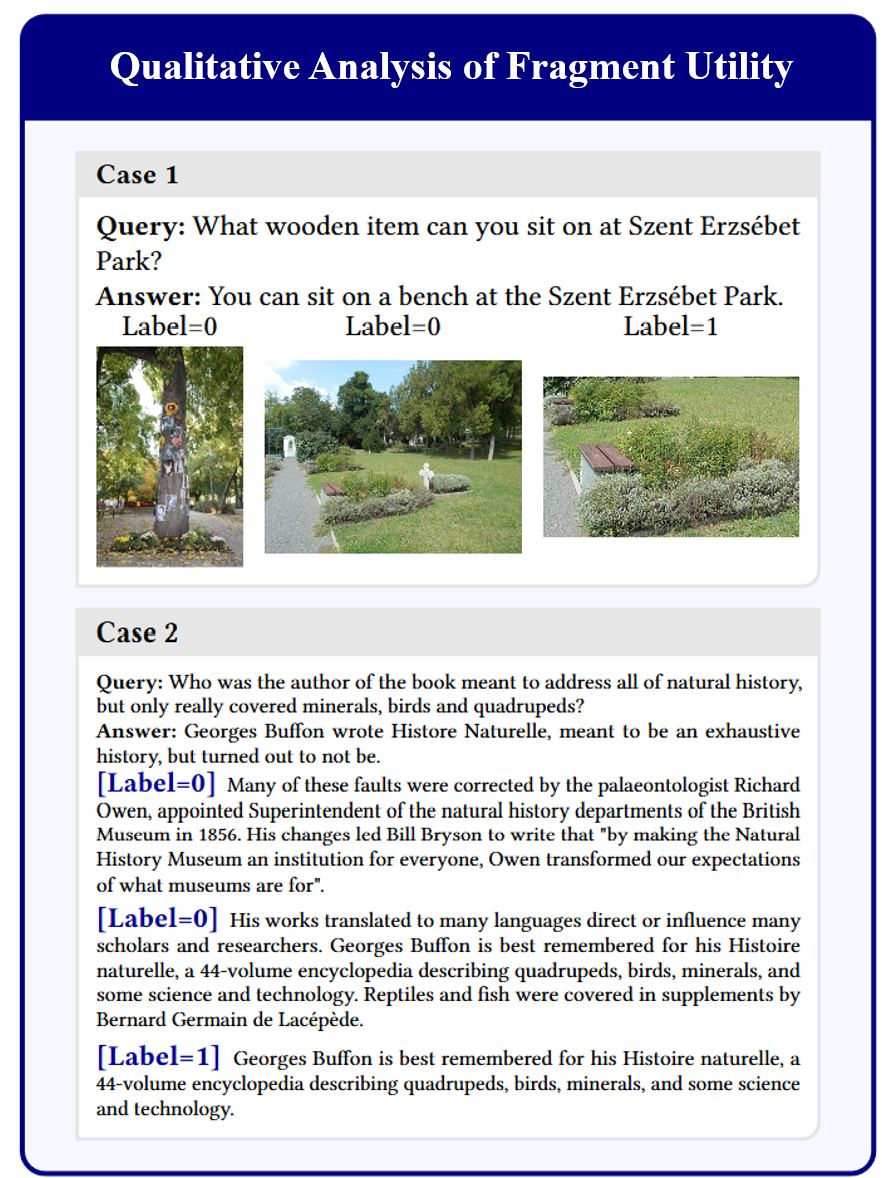} 
    % \caption{\textbf{Qualitative analysis of the FES-RAG framework.} \textbf{Figure A} demonstrates the fragment utility assessment, where the selector effectively filters out semantic noise (Label=0) to prioritize critical atomic evidence (Label=1) across both visual and textual modalities.}
    \caption{
Qualitative analysis of the FES-RAG framework. The selector effectively filters out semantically related but irrelevant fragments (Label=0) and prioritizes answer-critical evidence (Label=1) across both visual and textual modalities.
}
    \label{fig:qualitative_analysis}
\end{figure}

Figure \ref{fig:qualitative_analysis} illustrates the selector's ability to isolate utility-critical evidence. In \textbf{Case 1 (Visual)}, for the query ``wooden item you can sit on,'' it assigns low scores (Label=0) to semantically uninformative park scenery while accurately highlighting the \emph{bench} (Label=1). This demonstrates precise \emph{object-level semantic grounding} over coarse visual relevance. Similarly, in \textbf{Case 2 (Textual)}, the selector suppresses a keyword-rich but irrelevant distractor paragraph about ``Richard Owen'' (Label=0). Instead, it prioritizes the exact sentence identifying ``Georges Buffon'' (Label=1), exhibiting \emph{answer-bearing semantic precision} rather than relying on term frequency. 

Across modalities, FES-RAG consistently extracts atomic, answer-critical fragments. By filtering out semantically adjacent noise, it provides the MLLM with a compact, high-signal context, ensuring downstream reasoning is well-grounded and robust to distraction.

\section{Conclusion}
This paper presents FES-RAG, a framework that advances MRAG from monolithic document-level reranking to atomic fragment-level selection. It decomposes retrieved content into sentences and visual regions, enabling precise removal of redundant or noisy context that can impair MLLM reasoning. FES-RAG uses Fragment Information Gain (FIG) to estimate each fragment’s marginal utility, providing a more informative supervision signal than document-level relevance labels for training a fine-grained evidence selector. By distilling this utility assessment into a lightweight student model, FES-RAG achieves a 27\% relative CIDEr improvement on the M2RAG benchmark while substantially reducing token consumption. These results show that fragment-level purification effectively improves both factual grounding and inference efficiency in MLLMs.

\bibliographystyle{ACM-Reference-Format}
\bibliography{sample-base}

\end{document}